\documentclass[twocolumn,showpacs,amssymb,aps,prc,nofootinbib,floatfix,superscriptaddress]{revtex4-1}

\usepackage{epsfig}
\usepackage{graphicx}
\usepackage{float}
\usepackage{hyperref}
\usepackage{amsmath}
\usepackage{amssymb}

\begin{document}
	
\title{Cumulants of mean transverse momentum and elliptic flow in the hydrodynamic model of heavy-ion collisions}

\author{Tribhuban Parida}
\email{tparida@agh.edu.pl}
\affiliation{AGH University of Krakow, Faculty of Physics and
	Applied Computer Science, aleja Mickiewicza 30, 30-059 Cracow, Poland}

\author{Piotr Bożek}
\email{piotr.bozek@fis.agh.edu.pl}
\affiliation{AGH University of Krakow, Faculty of Physics and
	Applied Computer Science, aleja Mickiewicza 30, 30-059 Cracow, Poland}

\begin{abstract}
  Higher order cumulants between the mean transverse momentum and  elliptic flow are calculated in a relativistic viscous hydrodynamic model of relativistic heavy-ion collisions. The results of the hydrodynamic simulations are compared with calculations using  event-by-event predictors of the final collective observables  constructed from the initial state entropy distribution. The predictors  describes quantitatively centrality dependence of  the higher cumulants considered in the paper. We derive a quantitative relations between the cumulants of the mean transverse momentum and different  moments of the harmonic flow. The hydrodynamic simulations satisfy those relation very well. Those relations could be used to test experimentally the collective origin of the observed correlations between the mean transverse momentum and harmonic flow.
\end{abstract}


\maketitle

\section{Introduction}\label{introduction}

Many experimental and theoretical studies of relativistic heavy-ion collisions are based on the analysis of collective observables derived from the spectra of particles emitted in the collisions \cite{Ollitrault:2010tn,Heinz:2013th,Gale:2013da}. These observables involve event averages or moments of quantities such as the charged-particle multiplicity, the mean transverse momentum, and the harmonic flow coefficients of the azimuthal particle distribution. The variance, skewness, and kurtosis of the mean transverse momentum, as well as fluctuations of the radial flow, have been studied extensively \cite{Voloshin:1999yf,STAR:2003cbv,ALICE:2014gvd,NA49:2003hxt,Gavin:2003cb,Broniowski:2009fm,ALICE:2025rtg,ATLAS:2025ztg,Jia:2025rab,Parida:2024ckk,ATLAS:2024jvf}.

Cumulants of the harmonic flow coefficients have also been measured and calculated in models. The study of higher-order cumulants of harmonic flow coefficients is an important tool for identifying collective properties in heavy-ion collisions \cite{Borghini:2000sa,Borghini:2001vi,STAR:2002hbo,ALICE:2011ab,ATLAS:2014qxy,CMS:2013jlh,Bilandzic:2013kga}. Correlations between different flow harmonics have also been analyzed \cite{Teaney:2012ke,Aad:2014fla,ALICE:2016kpq,Qian:2016fpi,Zhu:2016puf,Gardim:2016nrr,Qiu:2012uy}.

Mixed correlations of the mean transverse momentum and harmonic flow have become a standard tool in the study of the collective expansion of the fireball formed in heavy-ion collisions \cite{Bozek:2016yoj,Schenke:2020uqq,ATLAS:2019pvn,Giacalone:2020awm,Giacalone:2020dln,Jia:2021wbq,Bally:2021qys,Fortier:2024yxs,Giacalone:2024luz,STAR:2024wgy,Li:2025hae,CMS:2024rvk}. Many-particle symmetric cumulants involving the mean transverse momentum and two or three different flow harmonics have been calculated in models and measured experimentally \cite{Bozek:2021zim,ALICE:2021gxt}. Measurements of cumulants involving the mean transverse momentum and the fourth power of harmonic flow have also been proposed \cite{CMS:2024rvk,Nielsen:2025pkz,Zhao:2024lpc}. Cumulants of higher powers of the mean transverse momentum have been investigated as well \cite{Nielsen:2025pkz,Zhao:2024lpc}.

The motivation for studying many-particle cumulants involving powers of the mean transverse momentum and harmonic flow is to gain sensitivity to many-body correlations in the initial-state density distribution. Such correlations may originate from quadrupole, octupole, or triaxial deformations, or from the presence of nuclear clustering in the colliding nuclei \cite{Giacalone:2023hwk,Mehrabpour:2026yuc,Zhao:2024lpc}.

In this paper, we investigate higher order cumulants  of the mean transverse momentum and  elliptic flow in a hydrodynamic model of heavy-ion collisions. The calculation of such higher-order correlators is numerically demanding. We study whether the scaled higher-order cumulants can be approximated using predictors based on event-by-event density profiles generated in the  model of the initial state chosen for the hydrodynamic evolution. We calculate the scaled cumulant of the mean transverse momentum with elliptic flow at order four or six order. Using a simple ansatz for the joint probability distribution of the mean transverse momentum and elliptic flow, we find a quantitative relations between cumulants of different order in harminc flow. Those relation explain previous results \cite{CMS:2024rvk,Nielsen:2025pkz} as well as our hydrodynamic simulations.

\section{Cumulants of mean transverse momentum and collective flow}

\label{sec:cumulants}

In this section we recall the formulas for the cumulants
of mean transverse momentum and harmonic
flow coefficients \cite{Nielsen:2025pkz},
based on the generating function approach.
Defining the harmonic flow coefficients in a heavy-ion  collision event from the  azimuthal angle distribution of emitted particles,
\begin{equation}
\frac{d^2N}{dp d\phi} = \frac{dN}{2\pi dp} \left(  1 + 2 \sum_{n=1}^{\infty} V_n(p)e^{i n \phi} \right) \ \ ,
\label{ptdist}
\end{equation}
the mean transverse momentum of particles created in that  event is defined as,
\begin{equation}
[p_T]=\frac {1}{N} \int_{p_{min}}^{p_{max}}    dp \ p  \frac{dN}{dp}
\end{equation}
and the momentum averaged harmonic flow coefficient is 
\begin{equation}
V_n =  \frac{1}{N}  \int_{p_{min}}^{p_{max}}  dp\   V_n(p) \frac{dN}{dp} \ ,
\end{equation}
where $N$ is the multiplicity in the event
\begin{equation}
N=\int_{p_{min}}^{p_{max}} dp\   \frac{dN}{dp} 
\end{equation}
and $V_n=v_n e^{i\Psi_n}$.

The usual correlation coefficient between the mean transverse momentum and the
harmonic flow coefficient is \cite{Bozek:2016yoj}
\begin{equation}
  \rho([p_T],v_n^2)= \frac{Cov([p_t],v_n^2)}{\sqrt{Var([p_T])Var(v_n^2)}} \ ,
  \label{eq:ptv}
  \end{equation}
where the covariance is
\begin{equation}
  Cov([p_T],v_n^2)=\langle [p_T] v_n^2 \rangle -\langle[p_t] \rangle \langle v_n^2\rangle \ ,
  \end{equation}
and the variances are
\begin{equation}
  Var([p_t])=\langle [p_T]^2 \rangle -\langle[p_t] \rangle^2=\sigma_{p_T}^2,
\end{equation}
and
\begin{equation}
  Var(v_n^2)=\langle \left(v_n^2 \right)^2\rangle -\langle v_n^2 \rangle^2 =\sigma_{v_n}^2 \ .
\end{equation}

Higher  moments or cumulants of the  mean transverse momentum and/or harmonic
flow coefficient can be derived from the corresponding generating functions
\cite{Borghini:2001vi}.
The generating function for the moments of the mean momentum $[p_T]$ and the harmonic flow coefficient $V_n$  can be defined as
\begin{equation}
  G[t,s,s^\star]=\sum_{i=0}^{\infty}\sum_{k=0}^{\infty} \langle [p_t]^i (V_n)^k (V_n^\star)^l \rangle \frac{t^i s^l s^{\star k}}{i! k! l!}  \ ,
\end{equation}
where $\langle \dots \rangle$ is the average over the ensemble of events.
The cumulant generating function is
\begin{equation}
  H[t,s,s^\star]=\ln \left( G[t,s,s^\star] \right)
\label{eq:cgf}
  \ .
\end{equation}

Using  deviations $\delta p_T= [p_T]- \langle[p_T] \rangle$ and $\delta v_n^2= v_n^2 -\langle v_n^2\rangle$ from the event average,
the cumulants are
\begin{equation}
  \langle [p_T] v_n^2 \rangle_c=\langle \delta [p_T] \delta v_n^2\rangle  = Cov([p_T],v_n^2) \ ,
  \end{equation}
\begin{equation}
  \langle [p_t]^2 v_n^2 \rangle_c= \frac{\partial^4 H[t,s,s^\star]}{\partial t^2 \partial s \partial s^\star}=
  \langle \delta [p_T]^2 \delta v_n^2 \rangle  \ ,
  \end{equation}
\begin{eqnarray}
  \langle [p_t]^3 v_n^2 \rangle_c &=&\frac{ \partial^5 H[t,s,s^\star]}{\partial t^3 \partial s \partial s^\star}=
  \langle \delta [p_T]^3 \delta v_n^2 \rangle   \nonumber \\ &-& 3 \langle \delta [p_T]^2  \rangle\langle \delta [p_t] \delta v_n^2  \rangle \ ,
  \end{eqnarray}
\begin{eqnarray}
  \langle [p_t]^4 v_n^2 \rangle_c &=&\frac{\partial^6 H[t,s,s^\star]}{\partial t^4 \partial s\partial s^\star}=
  \langle \delta [p_T]^4 \delta v_n^2 \rangle  \nonumber \\ &-& 6 \langle \delta [p_T]^2  \rangle\langle \delta [p_t]^2 \delta v_n^2 \rangle  \nonumber \\
  &-& 4\langle \delta [p_T]^3  \rangle \langle \delta [p_t] \delta v_n^2 \rangle  \ .
\end{eqnarray}

One can define the scaled cumulants \cite{Parida:2025eqv}
\begin{equation}
  \rho_{k,1}([p_T],v_n^2)=\frac{\langle [p_T]^k v_n^2 \rangle_c}{\sigma_{p_T}^k \langle v_n^2 \rangle} \ .
  \label{eq:rhos}
\end{equation}
The lowest order scaled cumulant
\begin{equation}
  \rho_{1,1}([p_T],v_n^2)= v_{02}
  \end{equation}
is the  correlation coefficient $v_{02}$ between the mean transverse momentum and harmonic flow defined in Ref. \cite{Parida:2025eqv}. The normalization differs from the correlation coefficient  $\rho([p_T],v_n^2)$ [Eq. (\ref{eq:ptv})] by replacing  $ Var(v_n^2)$ by $\langle v^2\rangle$. Both  normalizations involve flow cumulants at two-particle level, which may introduce some non-flow contribution.

The cumulants of fourth and sixth order in $V_n$ are 
\begin{eqnarray}
  \langle [p_T] V_n^2 V_n^{\star \ 2} \rangle_c &=&  \frac{\partial^5 H[t,s,s^\star]}{\partial t \partial s^2 \partial s^{\star\ 2}}=\langle [p_T] v_n^4 \rangle - 4 \langle [p_T] v_n^2\rangle \langle v_n^2\rangle  \nonumber \\
  &+&4 \langle [p_T]\rangle \langle v_n^2\rangle^2 - \langle [p_t] \rangle \langle v_n^4\rangle
\end{eqnarray}
and
\begin{eqnarray}
\langle [p_T] V_n^3 V_n^{\star\ 3} \rangle_c &=&   \frac{\partial^7 H[t,s,s^\star]}{\partial t \partial s^3 \partial s^{\star\ 3}} = \langle [p_T] v_n^6\rangle
-  \langle [p_t] \rangle \langle v_n^6\rangle \nonumber \\
  &-&   9 \langle [p_T] v_n^4 \rangle \langle v_n^2 \rangle 
+ 36 \langle [p_T] v_n^2 \rangle \langle v_n^2 \rangle^2
\nonumber \\ &-& 36  \langle [p_t] \rangle \langle v_n^2\rangle^3 
 - 9 \langle [p_T] v_n^2 \rangle \langle v_n^4 \rangle \nonumber \\
&+& 18  \langle [p_t] \rangle \langle v_n^2\rangle\langle v_n^4\rangle
  \end{eqnarray}

Using $\delta p_T $ and $\delta v_n^2$, one gets
\begin{equation}
  \langle [p_T] V_n^2 V_n^{\star\ 2} \rangle_c =\langle \delta p_T \delta v_n^2\delta v_n^2 \rangle -2 \langle v_n^2 \rangle \langle\delta p_T \delta v_n^2 \rangle \ 
\end{equation}
and
\begin{eqnarray}
  \langle [p_T] V_n^3 V_n^{\star\ 3} \rangle_c &=&\langle \delta p_T \delta v_n^2\delta v_n^2 \delta v_n^2 \rangle -6 \langle v_n^2 \rangle \langle\delta p_T \delta v_n^2 \delta v_n^2\rangle  \nonumber \\  &+&12  \langle \delta p_T \delta v_n^2 \rangle \langle v_n^2 \rangle^2 \nonumber \\ &-&9  \langle \delta p_T \delta v_n^2 \rangle \langle \delta v_n^2 \delta v_n^2 \rangle\ .
\end{eqnarray}
The corresponding scaled cumulants can be defined as \cite{Nielsen:2025pkz}
\begin{equation}
  \rho_{1,k}([p_T],v_n^2)=\frac{ \langle [p_T] V_n^k V_n^{\star\ k} \rangle_c }{\sigma_{p_T} c\{2k\}_n }\ .
  \label{eq:rho122}
  \end{equation}
where 
\begin{equation}
  c\{2k\}_n= \frac{\partial^{2k} H[t,s,s^\star]}{ \partial s^k \partial s^{\star\ k}}
  \end{equation}
is the $2k$ cumulant of the harmonic flow, $c\{2\}=\langle v^2 \rangle$,
$c\{4\}_n=\langle v_n^4 \rangle -2 \langle v_n^2\rangle^2 $, $c\{6\}_n=\langle v_n^6 \rangle +12 \langle v_n^2\rangle^3-9 \langle v_n^2\rangle  \langle v_n^4 \rangle $~.

\section{Hydrodynamic model calculations}

\label{sec:results}

We perform the relativistic hydrodynamic evolution of heavy-ion collisions using the \textsc{music} framework \cite{Schenke:2010nt,Schenke:2010rr,Paquet:2015lta}. The initial conditions for the hydrodynamic evolution are provided by the T$_{\rm R}$ENTo model \cite{Moreland:2014oya}, where the initial entropy density profile is parameterized as
\begin{equation}
s(x,y) = s_0 \sqrt{T_+(x,y)\, T_-(x,y)} \, ,
\end{equation}
with $T_{\pm}(x,y)$ denoting the thickness functions of the projectile and target participants, respectively. Centrality classification is performed using minimum-bias T$_{\rm R}$ENTo simulations, with the total initial entropy in the transverse plane used as an estimator of the final multiplicity.

\begin{figure}
	\vspace{5mm}
	\begin{center}
	  \includegraphics[width=0.4\textwidth]{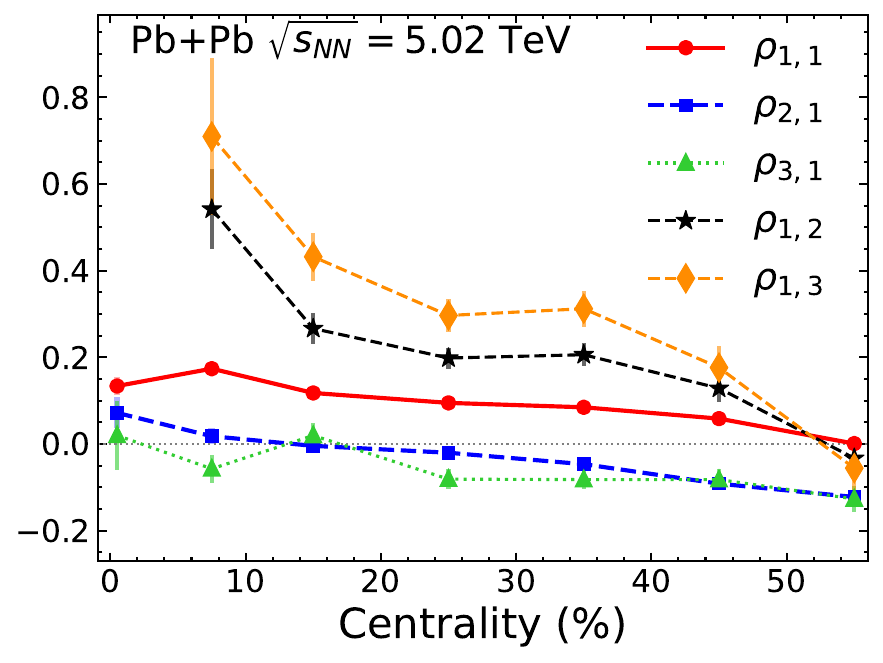} 
	\end{center}
	\caption{The scaled cumulants $\rho([p_T],v_2^2)$ (solid line),
          $\rho_{2,1}([p_T],v_2^2)$ (dashed line with squares), $\rho_{3,1}([p_T],v_2^2)$ (dotted line), $\rho_{1,2}([p_T],v_2^2)$ (dashed line with stars), $\rho_{13}([p_T],v_2^2)$ (dashed line with diamonds),  calculated in the relativistic hydrodynamic model for Pb+Pb collisions as a function of collision centrality.}
	\label{fig:rhos}
\end{figure}

The hydrodynamic evolution is performed independently for events in different centrality classes. The evolution starts at a constant proper time $\tau_0 = 0.4$ fm/$c$. During the evolution, we use a constant shear-viscosity-to-entropy-density ratio $\eta/s = 0.08$, together with a temperature-dependent bulk-viscosity profile taken from Ref.~\cite{Moreland:2018gsh}. We use the equation of state  from Ref.~\cite{Moreland:2015dvc}, based on lattice QCD at zero baryon chemical potential matched to a hadron resonance gas at low temperature. After the hydrodynamic evolution, freeze-out is performed on a constant-temperature hypersurface at $T_{\rm fr} = 145$ MeV using the Cooper--Frye prescription. Resonance decays are then included, and the resulting phase-space probability distributions of the final-state hadrons are used to calculate the hadronic observables.

The scaled cumulants $\rho_{k,1}([p_T],v_n^2)$ [Eq.~(\ref{eq:rhos})] and $\rho_{1,k}([p_T],v_n^2)$ [Eq.~(\ref{eq:rho122})] are calculated using events generated in the hydrodynamic model for Pb+Pb collisions at $\sqrt{s}_{NN}=5.02$~TeV. To correct for the effect of multiplicity fluctuations in each centrality bin, we correct the estimated deviations of the observables $O=[p_T]$ or $O=v_2^2$ from the event average \cite{Schenke:2020uqq},
\begin{equation}
\hat{\delta}O=\delta O - \frac{Cov(O,N)}{\sigma_N^2}\delta N \, .
\end{equation}
The cumulants are calculated using the observables corrected for multiplicity fluctuations, $\hat{\delta}O$. We have checked that, within statistical errors, these results are equivalent to those obtained by subdividing the centrality bins into several narrow $1\%$ bins and averaging.

The results in Fig.~\ref{fig:rhos} for the scaled cumulant $\rho_{1,1}([p_T],v_2^2)$ are consistent with other calculations and measurements \cite{ATLAS:2019pvn,Bozek:2020drh,Jia:2021wbq,Schenke:2020uqq}. The scaled cumulants of $[p_T]^2$ and $[p_T]^3$ are small in semicentral collisions, but become negative in peripheral collisions. This behavior is qualitatively consistent with results from the AMPT transport model shown in Ref.~\cite{Nielsen:2025pkz}. The scaled cumulants $\rho_{1,2}([p_T],v_2^2)$ and  $\rho_{1,3}([p_T],v_2^2)$ are approximately proportional and of the same sign as  $\rho_{1,1}([p_T],v_2^2)$. This is qualitatively consistent with preliminary results from the CMS Collaboration \cite{Tuo:2023tye} and with AMPT results \cite{Nielsen:2025pkz}. In Sec.~\ref{sec:GBG}, we derive, using a simple model, quantitative relations between the cumulants $\rho_{1,k}([p_T],v_2^2)$ ($k=1,2,3$).  Those relations explains their relative magnitudes and signs  found in the numerical simulations.

\section{Relation between the cumulant of the second and first order in $v_2^2$}

\label{sec:GBG}

Assuming a Gaussian distribution of eccentricity fluctuations around the geometric deformation of the interaction region, the event-by-event elliptic-flow distribution can be written as a Bessel--Gaussian distribution in the harmonic flow coefficient $v^2$ \cite{Voloshin:2007pc}:
\begin{equation}
  B(v^2)=e^{-\frac{v^2 + v_0^2}{2 s_v^2}} I_0\left( \frac{v_0\sqrt{v^2}}{s_v^2}\right) \ ,
  \label{eq:bg}
  \end{equation}
with
\begin{equation}
  \langle v^2 \rangle = v_0^2+2 s_v^2  \ ,
  \end{equation}
\begin{equation}
  c\{4\}= \langle v^{4} \rangle_c = - v_0^4 \ , 
\end{equation}
and
\begin{equation}
  c\{6\}= \langle v^{6} \rangle_c = 4 v_0^6 \ .
\end{equation}

Generalizing the distribution in Eq.~(\ref{eq:bg}) to a joint distribution of the mean transverse momentum $p$ and elliptic flow $v^2$, we take
\begin{equation}
P(p,v_2^2)= A \,
e^{-\frac{(p-\tilde{p})^2}{2 s_p^2}
+\frac{\kappa \left(v^2-v_0^2-2 s_v^2\right)(p-\tilde{p})}{s_p s_v^2}}
\, B(v^2) \, ,
\label{eq:GBG}
\end{equation}
assuming a Gaussian distribution in $p$, with a term in the exponent that correlates $p$ and $v^2$ with strength $\kappa$. The presence of correlations between the elliptic flow and the mean transverse momentum modifies the averages of the $v^2$ and $p$ variables.

The normalization coefficient is
\begin{equation}
A=\frac{1}{\sqrt{2 \pi} s_p}\left( 1+ \kappa^2 \left(2 +2 \frac{v_0^2}{s_v^2} \right)+ O(\kappa^4) \right) \, ,
\end{equation}
and the averages are modified at subleading order by the presence of correlations:
\begin{equation}
\langle p \rangle = \tilde{p}+ O(\kappa^3) \, ,
\end{equation}
\begin{equation}
\langle v^2 \rangle = v_0^2+2 s_v^2 +O(\kappa^2) \, ,
\end{equation}
\begin{equation}
c\{4\}^4=-v_0^4+O(\kappa^2) \, ,
\end{equation}
and
\begin{equation}
c\{6\}^4=4v_0^6+O(\kappa^2) \, .
\end{equation}

\begin{figure}
    \vspace{5mm}
    \begin{center}
        \includegraphics[width=0.4\textwidth]{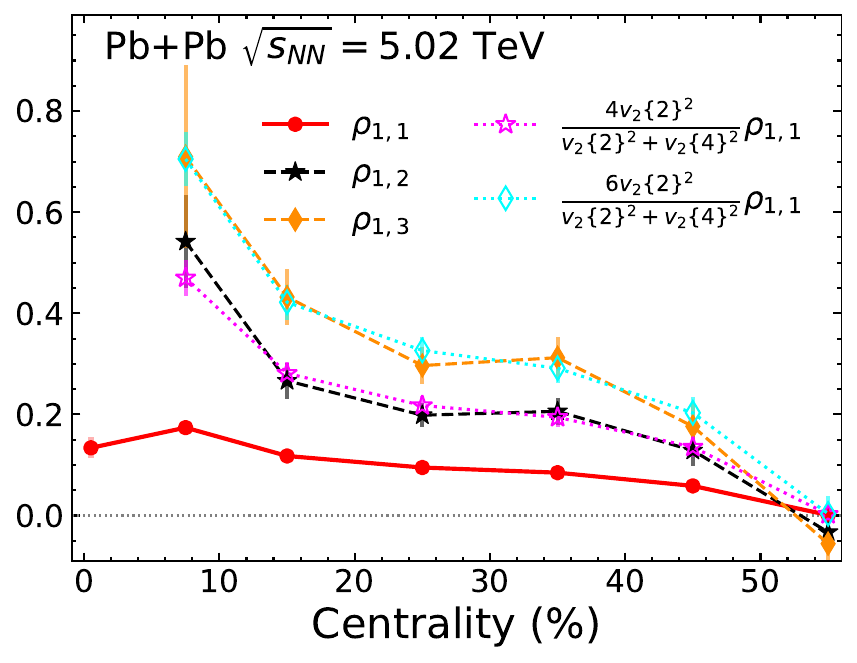}
    \end{center}
    \caption{Scaled cumulants $\rho_{1,1}([p_T],v_2^2)$ (solid line), $\rho_{1,2}([p_T],v_2^2)$ (dashed line with stars), and   $\rho_{1,3}([p_T],v_2^2)$ (dashed line with diamonds) calculated in the relativistic hydrodynamic model for Pb+Pb collisions as a function of collision centrality. The dotteded lines represent the estimates of $\rho_{1,2}$ and $\rho_{1,3 }$ obtained from $\rho_{1,1}$  using the scaling relations (\ref{eq:scaling}) and  (\ref{eq:scaling13}).}
    \label{fig:rho122}
\end{figure}

From the simple ansatz of a correlated $[p_T]$--$v_n^2$ distribution, Eq.~(\ref{eq:GBG}), one obtains to leading order in $\kappa$
\begin{eqnarray}
  \rho_{1,1}(p,v^2) & =&\frac{4 {s_v^2+v_0^2}}{2 s_v^2 +x_0^2} \kappa + O(\kappa^3)\nonumber \\
  &=&
    \frac{2 (\langle v^2\rangle + v\{4\}^2 )}{\langle v^2\rangle  }\kappa + O(\kappa^3)
\end{eqnarray}
\begin{equation}
  \rho_{1,2}(p,v^2)= 8  \kappa + O(\kappa^3)
\end{equation}
\begin{equation}
  \rho_{1,3}(p,v^2)= 12  \kappa + O(\kappa^3)
\end{equation}
To the lowest order in the correlation parameter $\kappa$, one finds
\begin{equation}
  \rho_{1,2}(p,v^2) \simeq  \frac{4 \langle v^2\rangle  }{ (\langle v^2\rangle + v\{4\}^2 )} \rho_{1,1}(p,v^2) + O(\kappa^2)
\label{eq:scaling}
\end{equation}
\begin{eqnarray}
  \rho_{1,3}(p,v^2) &\simeq&  \frac{6 \langle v^2\rangle  }{ (\langle v^2\rangle + v\{4\}^2 )} \rho_{1,1}(p,v^2) + O(\kappa^2) \nonumber \\ &\simeq& \frac{3}{2} \rho_{1,2}(p,v^2)+ O(\kappa^2) \ .
\label{eq:scaling13}
\end{eqnarray}
The parameterization in Eq.~(\ref{eq:GBG}) predicts that the scaled cumulants
$ \rho_{1,2}(p,v_2^2)$ and $ \rho_{1,3}(p,v_2^2)$  for the elliptic flow are of the same sign as the lowest order cumulant $\rho_{1,1}(p,v^2)$. The results in Fig. \ref{fig:rho122}
show that the relation Eq.~(\ref{eq:scaling}) is approximately fulfilled by the hydrodynamic model results. The ratios of the cumulants of different order in $v$ takes the form
\begin{equation}
\frac{\langle [p_T] V^2 V^{\star 2} \rangle_c}{\langle [p_T] v^2\rangle_c}= -\frac{4 v\{4\}^4 }{\langle v^2\rangle + v\{4\}^2}+O(\kappa^2) \, 
\end{equation}
and
\begin{equation}
\frac{\langle [p_T] V^3 V^{\star 3} \rangle_c}{\langle [p_T] v^2\rangle_c}= -\frac{24 v\{6\}^6 }{\langle v^2\rangle + v\{4\}^2}+O(\kappa^2) \, ,
\end{equation}
which could be tested experimentally as well.

The distribution in Eq.~(\ref{eq:GBG}) does not introduce independent parameters for the  higher order cumulants of the mean transverse momentum $[p_T]$  (skewness or kurtosis) or indpenedent correlations of the moments of $[p_T]$ with the harmonic flow. Therefore it is not expected to describe even qualitatively  the magnitudes of the scaled  cumulants $\rho_{2,1}(p,v^2)$ or   $\rho_{3,1}(p,v^2)$.
The ansatz in Eq.~(\ref{eq:GBG}) predicts for the cumulants of higher order in $[p_T]$
\begin{equation}
\rho_{2,1}(p,v^2)=\frac{4 (2 s_v^2+ 3 v_0^2)}{s_v \sqrt{s_v^2+v_0^2}} \kappa^2 + O(\kappa^4) \, ,
\end{equation}
and
\begin{equation}
\rho_{3,1}(p,v^2)=\frac{48 (s_v^2+ 2 v_0^2)}{s_v \sqrt{s_v^2+v_0^2}} \kappa^3 + O(\kappa^4) \, .
\end{equation}
These expressions reproduce neither the magnitude nor the sign of the hydrodynamic model results. The assumed distribution is Gaussian in $[p_T]$, with only a small correlation to $v^2$. A realistic description of the moments of $[p_T]$ and their correlation with $v_2^2$ therefore requires a more elaborate ansatz than Eq.~(\ref{eq:GBG}).

\section{Predictors of final correlations from the initial state}

\begin{figure*}
	\vspace{5mm}
	\begin{center}
	  \includegraphics[width=0.9\textwidth]{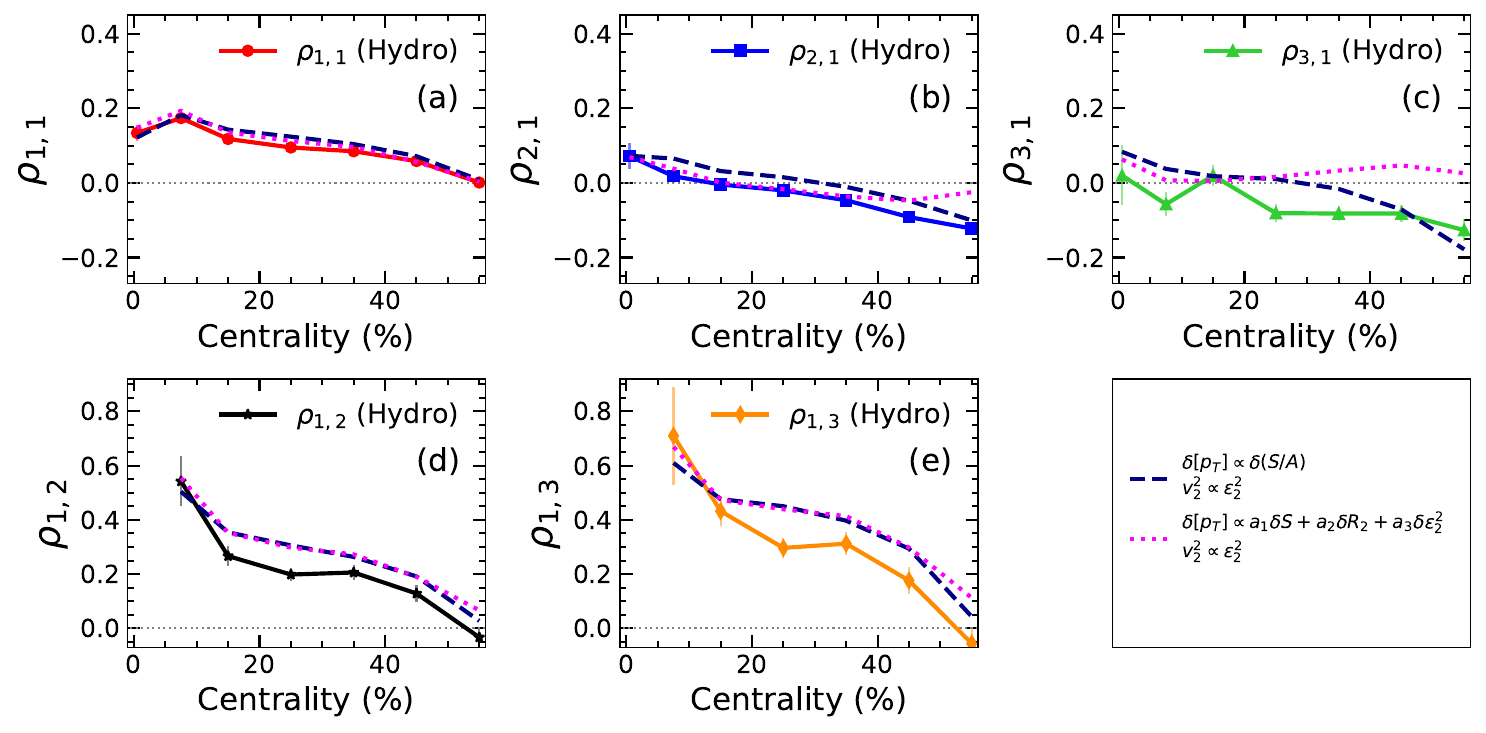} 
	\end{center}
	\caption{Comparison of the hydrodynamic-model results (solid lines with markers) with predictions based on estimators constructed from the initial density for the scaled cumulants of the mean transverse momentum and elliptic flow: $\rho_{1,1}([p_T],v_2^2)$ [panel (a)], $\rho_{2,1}([p_T],v_2^2)$ [panel (b)], $\rho_{3,1}([p_T],v_2^2)$ [panel (c)],  $\rho_{1,2}([p_T],v_2^2)$ [panel (d)], and $\rho_{1,3}([p_T],v_2^2)$ [panel (e)]. Estimates  using predictors  [Eq.~(\ref{eq:pred1})] and  [Eq.~(\ref{eq:pred2})] are shown with dashed, and dotted lines, respectively.}
	\label{fig:predictors}
\end{figure*}

It is instructive to examine the origin of the correlations that generate higher order cumulants between the mean transverse momentum and harmonic flow coefficients. In this section, we test how much of the correlations between the final-state observables can be explained using moments of the initial-state distribution \cite{Bozek:2020drh}. This requires constructing a predictor for the final observables based on properties of the initial state. Such a procedure is also important in practice, because the calculation of higher-order cumulants requires computationally demanding hydrodynamic simulations \cite{Schenke:2020uqq,Zhao:2024lpc,Bozek:2020drh,Bozek:2021zim}.



We use predictors for the final event-by-event observables $[p_T]$ and $v_2^2$, based on several characteristic moments of the initial-state entropy distribution in the transverse plane,
\begin{equation}
  [m]=\frac{\int m(x,y) s(x,y) dx dy }{\int s(x,y) dx dy} \ .
  \end{equation}
The moments of the initial density used are defined as 
\begin{equation}
  S=\int s(x,y) dx dy \ ,
\end{equation}
\begin{equation}
\epsilon_2^2= \left( \frac{[x^2-y^2]}{[x^2+y^2]}\right)^2 +  \left( \frac{[2xy]}{[x^2+y^2]}\right)^2\ ,
\end{equation}
\begin{equation}
  R_k= [(x^2+y^2)^{k/2}]^{1/k} \ ,
\end{equation}
and 
\begin{equation}
  A = \sqrt{[x^2][y^2]} \ .
\end{equation}

All predictors are applied to an independent sample of initial-state configurations with much larger statistics than the sample used for the hydrodynamic simulations. For predictors with many fitted parameters, using an independent sample of initial conditions reduces the effect of spurious overfitting to the simulated sample.

The first predictor \cite{Schenke:2020uqq} is defined as
\begin{eqnarray}
\delta [p_T] &\propto& \delta\left( \frac{S}{A} \right) \nonumber \\
v_2^2 &\propto& \epsilon_2^2 \, .
\label{eq:pred1}
\end{eqnarray}
This predictor does not require fitting any parameters to the results of the hydrodynamic simulations. However, there is no simple way to improve it systematically.

The  second predictor usees a linear approximations for the final observables based on moments of the initial density \cite{Bozek:2020drh}. The parameters of the linear predictor are fitted separately in each centrality bin to maximize the correlation between the prediction and the hydrodynamic result.
The mean transverse momentum is predicted using a combination of the lowest moments: the entropy $S$, the root-mean-square radius $R_2$, and the eccentricity $\epsilon_2^2$,
\begin{eqnarray}
  \delta [p_T] &\propto& a_1 \delta S + a_2 \delta R_2 + a_3 \delta \epsilon_2^2  \nonumber \\
   v_2^2 &\propto&  \epsilon_2^2  \ .
  \label{eq:pred2}
\end{eqnarray}

The predictions based on the initial density are compared with the hydrodynamic-model results in Fig.~\ref{fig:predictors}. None of the predictors describes all of the considered cumulants of the mean transverse momentum and elliptic flow well. The centrality dependence of the correlation coefficient $\rho([p_T], v_2^2)$ is reproduced qualitatively by all formulas [Fig.~\ref{fig:predictors}, panel (a)]. 
The hydrodynamic-model results for the cumulant $\rho_{2,1}([p_T], v_2^2)$ are also reproduced qualitatively the predictors [Fig.~\ref{fig:predictors}, panel (b)]. All of them show a change of sign as a function of centrality. However, the predictors do not reproduce well the hydrodynamic results for the cumulant $\rho_{3,1}([p_T], v_2^2)$ [Fig.~\ref{fig:predictors}, panel (c)]. Only the simplest predictor [Eq.~(\ref{eq:pred1})], shows a change of sign, but its centrality dependence is too steep compared with the hydrodynamic simulations. The centrality dependence of the fourth and six-order cumulants in elliptic flow, $\rho_{1,2}([p_T], v_2^2)$ and $\rho_{1,3}([p_T], v_2^2)$ is reproduced  qualitatively by the predictors [Fig.~\ref{fig:predictors}, panels (d) and (e)]. It is expected for any predictor reproducing the lowest oreder scaled cumulant  $\rho_{1,1}([p_T], v_2^2)$, due to the relations (\ref{eq:scaling}) and (\ref{eq:scaling13}).

The linear approximation for the mean transverse momentum can be improved further by including higher moments of the initial density \cite{Teaney:2010vd,Mazeliauskas:2015efa}~:
\begin{eqnarray}
\delta [p_T] &\propto& a_1 \delta S + a_2 \delta R_2 + a_3 \delta \epsilon_2^2 + a_4 \delta R_4 + a_5 \delta \epsilon_{4,2}^2 \nonumber \\
v_2^2 &\propto& \epsilon_2^2 \,  ,
\label{eq:pred3}
\end{eqnarray}
or
\begin{eqnarray}
  \delta [p_T] &\propto& a_1 \delta S + a_2 \delta R_2 + a_3 \delta \epsilon_2^2  +a_4 \delta R_4 + a_5 \delta \epsilon_{4,2}^2\nonumber \\
  \delta v_2^2 &\propto&  b_1 \delta S + b_2 \delta R_2 + b_3 \delta \epsilon_2^2  +b_4 \delta R_4 + b_5 \delta \epsilon_{4,2}^2 \ .
  \label{eq:pred4}
\end{eqnarray}
with 
\begin{equation}
\epsilon_{4,2}^2 = \left( \frac{[(x^4-y^4]}{[(x^2+y^2)^2]}\right)^2 +  \left( \frac{[2xy(x^2+y^2)]}{[(x^2+y^2)^2]}\right)^2\ .
\end{equation}
We have checked, that this improves slightly the correlation of the collective observables $[p_T]$ and $v_2^2$ with the predictors, but does not improve significantly the prediction for the scaled 
cumulants of the mean transverse momentum and elliptic flow.

\section{Conclusions}

Higher order correlation between the mean transverse momentum and the harmonic flow could be used to study many body correlations in the initial state related to the deformation or the nuclear cluster structure of colliding nuclei. First model calculations have been performed previously, but very few experimental results are available up to now.  In this paper higher order  cumulants of the elliptic flow and mean transverse momentum have been estimated for Pb+Pb 
collisions using a relativistic viscous hydrodynamic model.

With the available statistics, precise results can be obtained for the cumulants of second order in $[p_T]$, $\langle [p_T]^2 v_2^2 \rangle_c$, and up to six order in $v_2$, $\langle [p_T] V_2^3 V_2^{\star \ 3}\rangle_c$. However,  results for the higher cumulants  have larger statistical uncertainties. We find that the scaled cumulants of higher order in $[p_T]$ become negative at large centralities. 
We also investigate whether predictors of collective observables based on the event-by-event initial density distribution can be used to estimate the higher-order cumulants. We find that a simple predictor (\ref{eq:pred1}) describes   qualitatively   the centrality dependence of the scaledd cumulants considered here.

The scaled cumulants $\rho_{1,2}([p_T],v_2^2)$  and $\rho_{1,3}([p_T],v_2^2)$ are of the same order as lowest-order cumulant $\rho_{1,1}([p_T],v_2^2)$. Assuming a simple Gaussian form for the joint probability distribution of $[p_T]$ and $v_2^2$, we derive a simple relation between these the scaled cumulants of $[p_t]$ and higher orders of $v_2$. This relation is very well reproduced by the hydrodynamic-model results.

It would be interesting to test this relation in experimental data to confirm the collective-flow origin of the observed correlations. Measurements of the five-particle cumulants $\langle [p_T] V_2^2 V_2^{\star 2}\rangle_c$ and $\langle [p_T] V_2^3 V_2^{\star 3}\rangle_c$  could also be used to estimate possible nonflow effects in the lower-order cumulant $\langle [p_T] v_2^2 \rangle_c$ \cite{Zhang:2021phk}. 
Cumulants of higher order in the mean transverse momentum cannot be described by a simple  Gaussian ansatz. This makes the study of these higher-order cumulants interesting, because they contain information beyond that in the lowest-order correlation coefficient $\rho_{1,1}([p_T],v_2^2)$. The measurement of higher-order cumulants may also provide useful information in cases where the harmonic flow is dominated by fluctuations, such as in ultracentral nuclear collisions,  in proton--nucleus collisions or for the triangular flow.

\section*{Acknowledgments}
This research was supported by the Polish Ministry of Science and Higher Education  and
 by the  Polish National Science Centre Grant No. 2023/51/B/ST2/01625.

\section*{Data availability}

This manuscript has associated data openly available at \cite{Parida2026}.

\bibliographystyle{apsrev4-1}
\bibliography{references}

\end{document}